# Electron spin relaxation in intrinsic bulk InP semiconductor

Hong Ma, Zuanming Jin, Lihua Wang, Guohong Ma[a)]

*Department of Physics, Shanghai University, 99 Shangda Road, Shanghai 200444, P. R. China*

Electron spin dynamics in intrinsic bulk Indium Phosphide (InP) semiconductor is studied by time resolved pump probe reflectivity (TRPPR) technique using the co- and counter-circularly polarized femtosecond pulses at room temperature and 70 K. The reflectivity change from bleaching into absorption is observed with increasing pump photon energy, which can be explained in terms of the spin sensitive band filling and band gap renormalization effects. Density dependence of electron spin relaxation time shows similar tendency at room temperature and 70 K. With increasing carrier density, the electron spin relaxation time increases and then decreases after reaching a maximum value. Our experimental results agree well with the recent theoretical prediction [Jiang and Wu, Phys. Rev. B 79, 125206 (2009)] and D'yakonov-Perel' mechanism is considered as a dominating contribution to the electron spin relaxation in intrinsic bulk InP semiconductor.

[a)] Corresponding author. Electronic mail: ghma@staff.shu.edu.cn

## I. INTRODUCTION

In recent years, manipulation on spin dynamics in semiconductors and semiconductor nanostructures have attracted intense interest because of the potential applications in emerging areas such as “spintronics” and quantum information processing. A general understanding of the electron spin relaxation in semiconductor is a central topic for spintronics.[1-4] In fact, the spin relaxation is a complicated process because many parameters play important roles in determining the spin lifetime. [3,5] One example of the complication is carrier density dependence of spin relaxation in a semiconductor. The previous theory based on single particle approach predicts that spin relaxation time is monotonic donor concentration dependence.[6,7] However, recently, Wu et al. reexamined the electron spin dynamics in low-dimensional and bulk semiconductor in the metallic regime from a fully microscopic kinetic spin Bloch equation (KSBE) approach,[3,5,8-13] and they predicted a non-monotonic donor concentration and temperature dependence of the spin relaxation time, in which the maximum spin relaxation time occurs at the crossover of the degenerate regime to the nondegenerate one. [3,8,9] The theoretical prediction has been confirmed experimentally in n-type bulk GaAs and GaAs/AlGaAs quantum well,[14,15] and the experimental results can be well explained in term of the KSBE approach.[3,8,16] Among the spin relaxation study, intrinsic semiconductor is a good candidate for exploring scattering mechanism, because in intrinsic semiconductors electrons are generated together with equal number of holes by photoexcitation, the main scattering mechanisms are the carrier-carrier scattering and carrier-longitudinal-optical phonon scattering, and the carrier-impurity scattering rate is too small to contribute to the spin relaxation due to the absence of impurity. In addition, recent comprehensive theoretical investigation reveals that the carrier density dependence of spin relaxation in intrinsic bulk semiconductor shows a similar tendency with that in n-type semiconductor. [3,8] The interesting prediction is only partially confirmed experimentally up to now, [17-20] which leaves a lot of space to be verified further.

Indium phosphide, InP, with a similar band gap structure as that of well studied GaAs, has not been investigated systematically so far. Previous work reported that the electron spin relaxation time persists longer than 100 $\mu s$ in InP quantum dots. [21] By using magneto-optical Kerr effect, Kimel et al. [22,23] studied the dynamics of Kerr rotation and Kerr ellipticity in bulk InP near the spin-orbit split-off exciton transition at low temperature, a particular relaxation with time constant of 200 ps was reported, which was attributed to electron-spin relaxation process. Li *et al.*[24] studied the temperature dependence of electron spin dynamics in Fe-doped InP by utilizing time resolved Kerr rotation technique, and multi-exponential relaxation was observed. Besides magneto-optical Kerr technique, time resolved pump probe reflectivity (TRPPR) technique [18,25,26] with the co- and counter-circularly polarized pulses is a popular method, which has been proved to be one of the simple and powerful methods to characterize the degree of spin polarization and decay time in semiconductors, especially in thick and nontransparent samples.

Band gap energy $\mathrm{E_g}$ of bulk InP is 1.41 eV at 70 K and 1.34 eV at room temperature. Optical transitions take place around the Γ point of the first Brillouin zone. [7,18,27] An essential feature of the band structure is the fact that the top valence band is split into a degenerate $P_{3/2}$ band (heavy-hole states and light-hole states) and a spin-orbit split-off $P_{1/2}$ band that lies below $P_{3/2}$ with the energy difference of Δ=0.11 eV. In bulk semiconductor, the transitions excited by circularly polarized light, e.g., a left-handed circularly polarized $\sigma^-$, involves both heavy hole and light hole. The ratio of the optically generated spin-up $n\uparrow$ to spin-down $n\downarrow$ electrons is 3:1, owing to the form of the matrix elements for heavy-hole and light-hole inter-band transitions. Therefore, the maximum degree of net spin polarization is $P = (n\uparrow - n\downarrow)/(n\downarrow + n\uparrow) = 0.5$. [7]

In this paper, the electron spin dynamics in intrinsic bulk InP at 70 K and room temperature is studied by using femtosecond TRPPR technique. The experimental results reveal that circularly-polarized pump-induced reflectivity change of probe beam shows a positive value (photo-bleaching) when photon energy is slightly higher than InP band gap energy, while a negative change (photo-absorption) occurs when photon energy is much higher than the band gap energy. A model by taking account of spin sensitive band filling (BF) and band gap renormalization (BGR) effects is proposed to explain these phenomena successfully. In addition, it is observed that the electron spin relaxation time ($\tau_s$) shows a non-monotonic dependence on the carrier density at both low and room temperature: with increasing photoexcitation, $\tau_s$ increases and then decreases after reaching a maximum value of 2.1 ns at 70 K, and 52 ps at room temperature. This non-monotonic density dependence of spin relaxation is consistent with the theoretical prediction based on a microscopic KSBE approach proposed by Jiang and Wu.[8]

## Ⅱ.EXPERIMENTS

The InP single crystal (purchased from Hefei Kejing Materials Technology co., LTD) was grown by Bridgman method with crystalline orientation along [100] direction and the thickness is 500 μm. The sample was not intentionally doped with the background doping level of $9\times10^{14}$ /cm$^3$. The measurement of the transient reflectivity change of probe pulse was carried out by using the TRPPR technique. Both pump and probe pulses were generated by a tunable mode-locked Ti: Sapphire laser with a pulse width of 100 fs and a repetition rate of 80 MHz (Mai

Tai HP 1020, Spectra-Physics). The two beams with the intensity ratio of more than 10:1 were focused on the same spot of the sample. The pump beam reflected by the sample was blocked with a non-transparent aperture, while the probe beam was guided into a photodiode connected with a lock-in amplifier. The polarizations of both beams were adjusted independently by two achromatic quarter wave plates. The sample was mounted on a cold finger in a closed cycle liquid-He cooled optical cryostat with four transparent windows.

For a given pump pulse, the excited carrier density without considering two-photon absorption can be estimated by $n = (1-R)E\alpha/(Ah\nu)$,[28] $R$ is the reflectivity without pump pulse, $E$ is the photon energy per pulse, $A$ is the area of focused spot, $\alpha$ is the absorption coefficient of the sample at the wavelength of pump pulse. The photon energy $E$ is adjusted to be larger than band gap energy, but less than the spin-orbit split-off energy in all the experiments.

## Ⅲ. RESULTS AND DISCUSSION

Figure 1 shows the experimental data measured at room temperature with two representative photon energies of 1.393 eV and 1.425 eV, respectively. With photon energy of 1.393 eV and photoexcitation density of $1.18\times10^{17}$ /cm$^3$, Fig. 1 (a) shows a dramatic rise in reflectivity change followed by a long recovery process. The curves labeled as ($\sigma^+,\sigma^+$) and ($\sigma^+,\sigma^-$) are denoted as co- and counter-circularly polarized pump and probe beams, respectively. It is also seen that the amplitude of the curve ($\sigma^+,\sigma^+$) is much stronger than that of the curve ($\sigma^+,\sigma^-$) around zero delay time. Finally both ($\sigma^+,\sigma^+$) and ($\sigma^+,\sigma^-$) curves tend to completely coincide with each other, which indicates the spin polarization relaxation between |+1/2> and |-1/2> spin states in conduction band. The spin relaxation time can be obtained by the difference between ($\sigma^+,\sigma^+$)

and ($\sigma^+,\sigma^-$), which is shown in the inset of the Fig. 1(a). The difference can be well reproduced using a single exponential decay. Recalling that $\tau_s / 2 = \tau_{fit}$,[26,29] the spin relaxation with time constant of 50±1 ps is obtained for InP semiconductor according to our experimental data. Similar measurements were carried out with photon energies of 1.425 eV, much higher than the band gap energy. Compared with the photoexcitation at 1.393 eV, it is seen that the saturation amplitude of ($\sigma^+,\sigma^+$) around zero delay time is smaller by a factor of 2, and the following recovery signal even evolves into an absorption after about 50 ps as shown in Fig. 1(b). It is also seen that the curve ($\sigma^+,\sigma^-$) even shows negative reflectivity change around zero delay time, and the signal doesn't show tendency of going up in the measured range. By utilizing the same method mentioned above, the spin relaxation time is extracted to be 55.2±0.4 ps.

Figure 2 shows the time dependent reflectivity response of co- and counter-circularly polarized pump and probe pulses with three typical photon energies at 70 K. The time evolution reflectivity change with photon energy of 1.425 eV (Fig. 2(a)) resembles that of 1.393eV (Fig. 1(a)) at room temperature. The difference of two curves between ($\sigma^+,\sigma^+$) and ($\sigma^+,\sigma^-$) can be well fitted with biexponential decay with time constant of $\tau_{fit1}$=65 ps, $\tau_{fit2}$=$\tau_s$/2=470 ps which is presented in the inset of Fig. 2(a). This biexponential behavior at high photoexcitation and low temperature can be understood as: under high nonequilibrium distribution of the electrons, spin relaxation is accompanied by the initial thermalization and followed by cooling processes. Thus the relaxation is a complicated process which normally cannot be described by single exponential decay. If the photoexcitation density is very high, carrier-carrier collisions thermalize the carriers to hot plasma before emitting a large number of phonons. This hot plasma cools to the lattice temperature primarily by emitting optical phonons.[30] At low carrier density or high temperature (room temperature), the carrier cooling process is very fast and the corresponding amplitude is very small compared with the spin relaxation. And a single exponential spin relaxation is seen at high temperature (Fig. 1) and low photoexcitation (Fig. 2(c)). Therefore, we infer that the fast component at low temperature comes from the cooling of hot carriers and the slow component comes from electron spin relaxation. By increasing the photon energy to 1.512 eV (Fig. 2 (c)), the reflectivity change around zero delay time shows a negative value for both ($\sigma^+,\sigma^+$) and ($\sigma^+,\sigma^-$). It is noted that the absolute amplitude of the curve ($\sigma^+,\sigma^-$) is stronger than that of the curve ($\sigma^+,\sigma^+$), and the electron spin relaxation time is calculated to be 707±1 ps, which is shorter than that at 1.425 eV. When the photon energy is tuned to 1.494 eV, photo-bleaching for ($\sigma^+,\sigma^+$) and photo-absorption for ($\sigma^+,\sigma^-$) are observed, which are presented in Fig. 2 (b).

Sign changes in pump probe transmission measurements have been reported previously. Lai *et al.* [31] have reported that the saturation peak evolutes to an absorption enhancement as photon energy is larger than 1.56 eV in intrinsic GaAs at 9.6 K with magnetic field of B=1.3 T. A reversal sign of wavelength dependence of Kerr rotation curve was also observed at low temperature. [22-24] Generally, the reflectivity change of probe beam can be attributed to the effects of the carriers generated by the pump pulses via inter-band transitions, [32,33] which is related to the changes in absorption coefficient and refractive index of the semiconductor induced by pump beam. Therefore, the reflectivity change is very sensitive to the photoexcited carrier density. It is known that the states in conduction band are occupied by electrons and the states in valence band are empty after photoexcitation, namely, band filling (BF) effect. On the other hand, the band gap energy is also affected by band gap renormalization (BGR). The electrons with the same spin will repel and avoid one another. The screening of electrons in conduction band leads to a decrease in their energy, and lowering the energy of the bottom conduction band. The same also holds for the holes, which results in an increase of the energy of the top valence band. Therefore, BGR effect induces the band gap shrinking. BF and BGR effects show different photon energy and carrier density dependence, and they compete with each other in determining the magnitude and signature of the total PPR signal.

In order to elucidate the photon energy and carrier density dependence of the sign change in pump-probe reflection measurement, the photoinduced reflectivity changes considering both BF and BGR effects are calculated as follows. If parabolic band is assumed, the absorption coefficient is given by the square-root law: $\alpha_0(E,N)=\sum_{\nu} C_{\nu}\sqrt{E-E_g(N)}/E$, where $\nu$ =*hh, lh*, and $C_{hh}$ and $C_{lh}$ are the constants of the heavy and light hole related effective mass, *E* is the photon energy. The curve ($\sigma^+,\sigma^+$) reflects the decay of majority spin population (spin down$\downarrow$) after photoexcitation with $\sigma^+$ pump pulses, while the curve ($\sigma^+,\sigma^-$) reflects the increase of minority spin state (spin down$\uparrow$) from the majority spin state. After taking BF effect into account, the spin-related absorption coefficient can be modified as $\alpha_0^{\pm}(E,N^{\pm})(f_v(E,N^{\pm})-f_c(E,N^{\pm}))$ [32]. $f_c(E,N^{\pm})$ and $f_v(E,N^{\pm})$ are the probabilities of the conductor and valence band state being occupied by electrons, which are given by the Fermi-Dirac distribution functions. In addition, the renormalized band gap energy $E_{g'}(N^{\pm})$ also affects the occupancies, which is determined by the band gap shrinkage $\Delta E_g(N^{\pm})$.[32] After considering both BF and BGR effects, the modified absorption coefficient is given by $\alpha_0^{\pm}(E,E_{g'}(N^{\pm}))(f_v(E,N^{\pm})-f_c(E,N^{\pm}))$. The spin-related refractive index $n(E,N^{\pm})$ is obtained by Kramers-Krönig (K-K) relation based on absorption coefficient. The total reflectivity of probe beam is followed by $R=((n-1)^2+\kappa^2)/((n+1)^2+\kappa^2)$, $\kappa$ is the extinction coefficient related the absorption coefficient by $\kappa = \lambda / 4\pi\alpha$. Considering the effects mentioned above, we can calculate the change in the reflectivity $\Delta R$, which is the difference between the reflectivity $R'$ with the pump pulse and $R$ without pump pulse. It should be mentioned that carrier diffusion effect is not included because the calculated reflectivity change are only performed around the zero delay time. In addition, with carrier density below $5\times10^{17}$ /cm$^3$, free carrier absorption is much less important than BF and BGR effects, and it is also excluded during calculation.

Figure 3(a) presents the calculated results by taking account of the spin related BF and BGR effects with photoexcitation density of $1\times10^{17}$ /cm$^3$ at 70 K. The solid and dash lines are calculated from co- and counter-circularly polarized pump and probe beams, respectively. Near the band gap, the signal variation with energy is quite distinct in magnitude due to the band edge effects. Another remarkable feature is that the reflectivity change shows a large positive peak around the renormalized band gap energy, which is due to the BGR effect, while the negative reflectivity change appearing around the original band gap energy is dominated by BF effect. From the inset of Fig. 3(a), it is clearly seen that the total change in reflectivity is positive when photon energy is slightly larger than band gap energy, while it undergoes a negative sign with much larger photon energy. The critical photon energy (in which the reflectivity change undergoes from positive to negative) is quite different for ($\sigma^+$, $\sigma^-$) and ($\sigma^+$, $\sigma^+$) excitation. Fig. 3(b) shows the calculated co-circularly polarized pump and probe reflectivity changes at 70 K with carrier densities of $4\times10^{16}$ /cm$^3$, $6\times10^{16}$ /cm$^3$, $1\times10^{17}$ /cm$^3$, and $2\times10^{17}$ /cm$^3$, respectively. It is seen that both the total reflectivity change and the critical photon energy are sensitive function of the carrier density. The similar results are obtained for counter-circularly polarized case (not present here). To conclude, the sign of reflectivity changes in probe beam is well explained by spin-related BF and BGR effects, which shows both carrier density and photon energy dependence.

In order to further understand the mechanisms of the electron spin relaxation process, the dependence of spin relaxation time on carrier density is investigated systematically. With fixed photon energy, at 1.393 eV and 1.425 eV respectively, Fig. 4 summarizes the results of carrier density dependence of spin relaxation time at room temperature (square) and 70 K (triangle). The most interesting feature in Fig. 4 is that the spin relaxation time shows a non-monotonic carrier density dependence: at room temperature with carrier density below $2\times10^{17}$ /cm$^3$, $\tau_s$ increases with carrier density, when density is higher than $2\times10^{17}$ /cm$^3$, $\tau_s$ shows a decrease with excitation density. And for the case of 70 K, the maximum value of $\tau_s$ is observed at carrier density of ~$1.3\times10^{17}$ /cm$^3$. The nonmonotonous carrier density dependence of spin relaxation time is different from the previous report in an intrinsic GaAs, [34]in which spin relaxation time is observed to increase monotonously with the carrier density.

For bulk III-V semiconductors in metallic regime, the spin relaxation mechanisms are the D'yakonov-Perel'(DP)[35], the Elliott-Yafet (EY)[36,37] and the Bir-Aronov-Pikus (BAP)[38] mechanisms. Previous theoretical calculations and experimental studies indicated that the EY mechanism is unimportant for electron spin relaxation in an intrinsic semiconductor.[7,34] The single particle theory based on elastic scattering approximation indicated that BAP mechanism usually dominates the spin relaxation at low temperature and high hole density.[7] Recently, Zhou and Wu have found that the previous BAP mechanism based on single particle approach largely overestimated the spin relaxation rate. [10] The reason is that the previous BAP theory neglected the contributions of Pauli blocking to spin relaxation rate.[3, 10] And They came to conclude that BAP mechanism is unimportant for spin relaxation in almost all intrinsic bulk III-V semiconductors, and spin relaxation time is mainly dominated by DP mechanism even at low temperature. Based on elastic scattering approximation, the spin relaxation time due to the DP mechanism is [7]

$$\tau_s^{DP-1} = \frac{Q\alpha_c^{\ 2}(k_B T)^3}{\hbar^2 E_g}\tau_p \qquad (1)$$

Where Q is a parameter related with the scattering process, $\alpha_c$ is determined by $\alpha_c = 4m_e\eta(1-1/3\eta)^{-1/2}/(3m_{cv})$ and $\eta = \Delta/(\Delta + E_g)$. $\tau_p$ is the momentum relaxation time. According to this analytic formula, the spin relaxation time due to the DP mechanism is expected to be monotonic carrier density dependence. Jiang and Wu also pointed out that the DP mechanism plays a dominant role for electron spin relaxation process even at high carrier density and low temperature for both n-type and intrinsic semiconductor. [3,8] This conclusion is based on a microscopic KSBE approach they developed. This many-body approach includes all the relevant scatterings such as electron-electron scattering, electron-impurity scattering, electron-phonon scattering, electron-hole Coulomb scattering, and electron-hole exchange scattering explicitly. According to the KSBE approach, both the EY and BAP mechanisms are not important, [8,10] and the DP mechanism plays a dominant role for electron spin relaxation process in most of bulk semiconductors. Non-monotonic density dependence of spin relaxation time is one of the main predictions of the KSBE approach. [3,8]

In an intrinsic semiconductor with very low impurity density, electron-impurity scattering rate is negligible. And the electron-hole Coulomb scattering term can also be eliminated due to the short lifetime of hole spin. The spin relaxation time are mainly dominated by inhomogenous broadening, the electron-electron Coulomb scattering and the electron-LO phonon scattering (dominate at high temperature). After considering many-body effects, the modified traditional DP mechanism based on single particle theory (Eq. 1) can be employed to explain our experimental data qualitatively. The modified Eq.1 is given as [8, 39]

$$\tau_s^{DP-1} = \left\langle \left( |\Omega(k)|^2 - \Omega_z^2(k) \right) \right\rangle \tau_p^* \qquad (2)$$

The term, $\left\langle \left( |\Omega(k)|^2 - \Omega_z^2(k) \right) \right\rangle$, is the inhomogenous broadening of k-dependent frequency of spin precession. $\tau_p^*$ is the modified momentum relaxation time including the electron-electron ($\tau_p^{ee}$) scattering and electron-phonon ($\tau_p^{ep}$) scattering by the relation of $1/\tau_p^* = 1/\tau_p^{ee} + 1/\tau_p^{ep}$. Obviously, spin relaxation time is dominated by the two terms: the inhomogenous broadening and the modified momentum relaxation time. And the total spin relaxation time is $\tau_s = \tau_s^{ee} + \tau_s^{ep}$, with $\tau_s^{ee}$ and $\tau_s^{ep}$ denote for spin relaxation time due to the electron-electron and electron-phonon scatterings, respectively. The inhomogeneous broadenings in nondegenerate (low density) and degenerate (high density) regimes are approximated by [3,8]

$$\left\langle |\Omega(\mathrm{k})|^2 - \Omega_z^2(\mathrm{k}) \right\rangle \sim \mathrm{n_e^0 T^3} \qquad \text{for T} \gg \mathrm{T_F} \qquad (3a)$$

$$\left\langle |\Omega(\mathrm{k})|^2 - \Omega_z^2(\mathrm{k}) \right\rangle \sim \mathrm{n_e^2 T^0} \qquad \text{for T} \ll \mathrm{T_F} \qquad (3b)$$

Here, $T_F$ stands for Fermi temperature.

The modified momentum relaxation time, $\tau_p^*$, are mainly contributed by electron-electron scattering and electron-phonon scattering in an intrinsic semiconductor. The electron-electron scattering time is estimated by [8, 40]

$$\tau_p^{ee} \sim n_e^{-1} T^{3/2} \qquad \text{for T} \gg \mathrm{T_F} \qquad (4a)$$

$$\tau_p^{ee} \sim n_e^{2/3} T^{-2} \qquad \text{for T} \ll \mathrm{T_F} \qquad (4b)$$

Based on the Eqs.2-4, it can be obtained $\tau_s^{ee} \sim n_e T^{-9/2}$ ($T \gg T_F$) in nondegenerate regime, and $\tau_s^{ee} \sim n_e^{-8/3} T^2$ ($T \ll T_F$) in degenerate regime, respectively. At low photoexcitation density, the electron system is nondegenerate; the inhomogeneous broadening is insensitive to electron density and $\tau_p^{ee}$ decreases with electron density according to Eq. (4a). As a result, the spin relaxation time ($\tau_s^{ee}$) increases with electron density. In high carrier density regime, carrier density dependence of inhomogeneous broadening increases more efficient than that of the electron-electron scattering, as a result, the spin relaxation time decreases with carrier density in the degenerate regime. The crossover between the degenerate and nondegenerate can be estimated by the Fermi temperature, $T_F$. [8]Therefore, it is expected that density dependence of spin relaxation time is non-monotonic and exhibits a peak at particular carrier density, $n_c$. In addition to the electron-electron scattering, the electron-LO phonon scattering also plays an important role in the spin relaxation time at room temperature. [3]Previous calculations indicated that the electron-LO phonon scattering time $\tau_p^{ep}$ is almost carrier density independent in low density (nondegenerate) regime, whereas $\tau_p^{ep}$ increases with carrier density in high density (degenerate) regime. [3]Therefore, in low density regime, spin relaxation time $\tau_s^{ep}$ due to the electron-phonon scattering changes very little with carrier density and it decreases with density in high carrier density regime. To sum up, after including the contributions from both electron-electron scattering and electron-phonon scattering, the total spin relaxation time $\tau_s$ is predicted to show a non-monotonic carrier density dependence and a peak spin relaxation time is expected to appear at the crossover of the degenerate regime to the nondegenerate one, which is due to the inelastic electron-electron and electron-phonon scattering and the influence of screening.

The nonmonotonous carrier density dependence of spin relaxation time observed in intrinsic InP semiconductor can be interpreted successfully with the modified DP mechanism mentioned above. [8]In an intrinsic semiconductor, the nonmonotonic density dependence of the spin relaxation time is mainly dominated by the electron-electron scattering and electron-phonon scattering, and the peak density $n_c$ appearing in the crossover regime arises from the competing effect between density dependence of inhomogeneous broadening and the density dependence of momentum electron-electron scattering and electron-phonon scattering. The crossover regime corresponds to a transformed density region, where carrier density lies from nondegenerate to degenerate regimes.

It is also seen from Fig. 4 that the carrier density at peak spin relaxation time at 70 K (~$1.3\times10^{17}$ /cm$^3$) is smaller than that (~ $2\times10^{17}$ /cm$^3$) at room temperature, which shows agreement with the theoretical expectation in GaAs qualitatively. [3,8] At low temperature, spin relaxation time is insensitive to electron-LO phonon scattering and the electron-electron scattering play dominating role for the case of carrier density dependence of spin relaxation. The critical carrier density from nondegenerate to degenerate region is mainly dependent on the electron-electron scattering rate. At high temperature, the electron-LO-phonon scattering, besides the electron-electron scattering, plays a significant role to electron spin relaxation, but it has different carrier density dependence. The rate of the electron-electron scattering increases much faster than that of the electron-LO-phonon scattering as the carrier density increases. Therefore, a higher carrier density at high temperature is needed than that at low temperature for the regime to enter into a degenerate regime.

## IV. Summary

In conclusion, electron spin relaxation in InP crystal is investigated by TRPPR technique at room temperature and 70 K. The reflectivity signals of probe beam experience from photo-bleaching to photo-absorption with increasing the photon energy. The change of reflectivity sign is interpreted by spin dependent band filling and band gap renormalization effect. By varying carrier density, it is found that the spin relaxation time increases, reaching a maximum value of ~2.1 ns (52 ps) at carrier density of $1.3\times10^{17}$ /cm$^3$ ($2\times10^{17}$ /cm$^3$) at 70 K (room temperature), then it shows continuously decrease with carrier density. The nonmonotonous dependence of spin relaxation on carrier density can be well explained by DP mechanism based on KSBE approach.

## Acknowledgements

This work is supported by National Natural Science Foundation of China (10774099), the Program for Professor of a Special Appointment (Eastern Scholar) at Shanghai Institutions of Higher Learning, and Science and Technology Commission of Shanghai Municipal (09530501100). Part of the work is also supported by Shanghai Leading Academic Discipline Project (S30105)

**Captions**

FIG. 1. Pump-induced probe reflectivity change for co-circularly polarized pulse ($\sigma^+, \sigma^+$) and counter-circularly polarized pulse ($\sigma^+, \sigma^-$) as a function of delay time at room temperature. Pump photon energy is tuned at 1.393eV (a) and 1.425eV (b), respectively, which are larger than the band gap energy, but less than spin-orbit split-off energy.

FIG. 2. Time evolution of reflectivity changes of probe beam at 70 K. Pump photon energies are 1.425eV (a), 1.494eV (b) and 1.512eV (c).

FIG. 3. (a) The solid and dash lines are calculated from co-circularly and counter-circularly polarized pump and probe beams, respectively. The positive and negative peaks are at corresponding to the renormalized band gap energy and the original band gap energy, respectively. The arrows in the inset indicate the critical photon energy transforming from bleaching to absorption. (b) The calculated reflectivity changes at 70K with carrier densities of $4\times10^{16}$ /cm$^3$ (solid line), $6\times10^{16}$ /cm$^3$ (dash line), $1\times10^{17}$ /cm$^3$ (dot line), and $2\times10^{17}$ /cm$^3$ (dash dot line) from co-circularly polarized pump and probe. The inset is the enlargement of the shading area.

FIG. 4. Carrier density dependence of spin relaxation time constant with fixed pump photon energy of 1.425 eV at 70K (triangle) and 1.393 eV at room temperature (square).

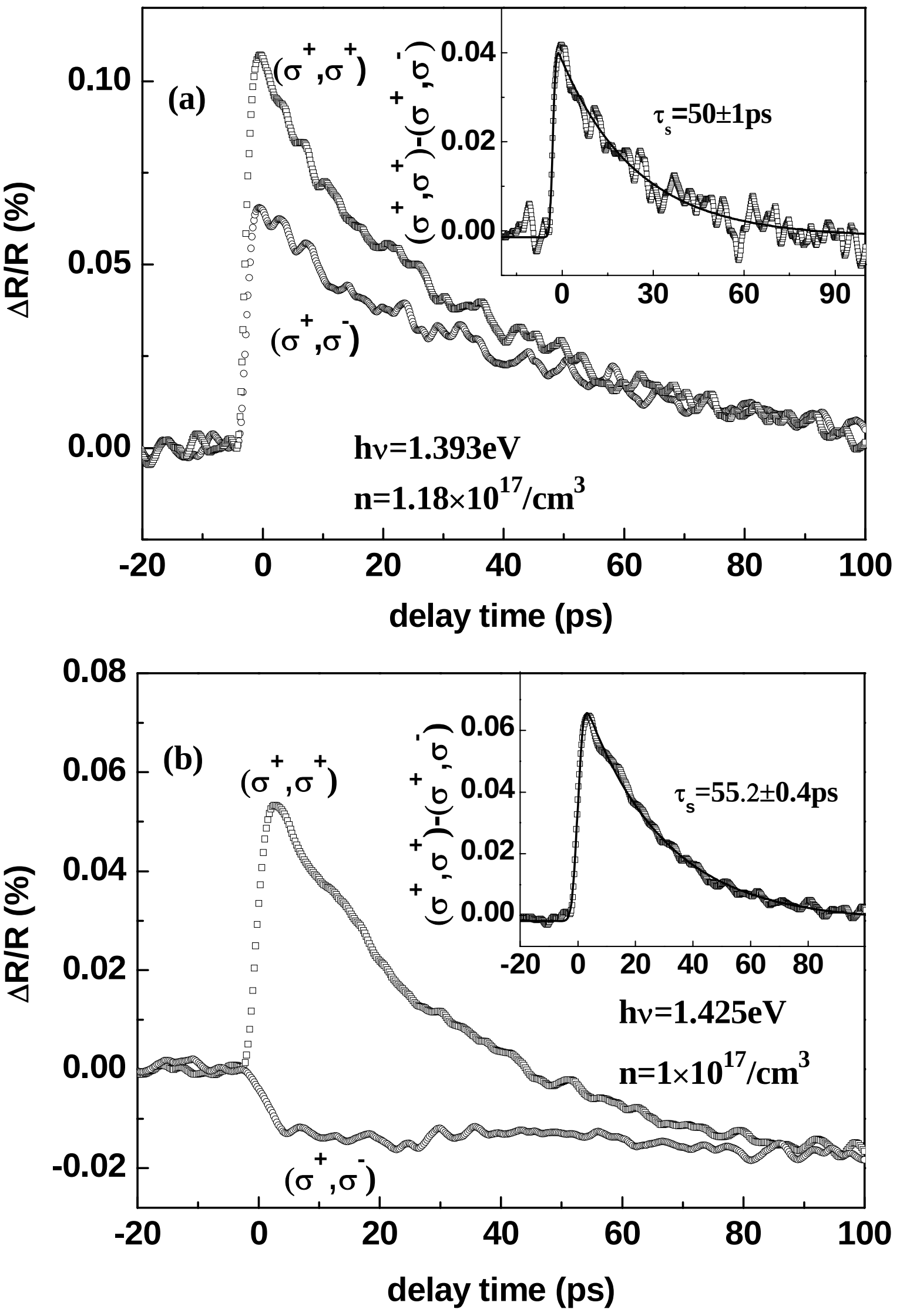

Fig. 1.

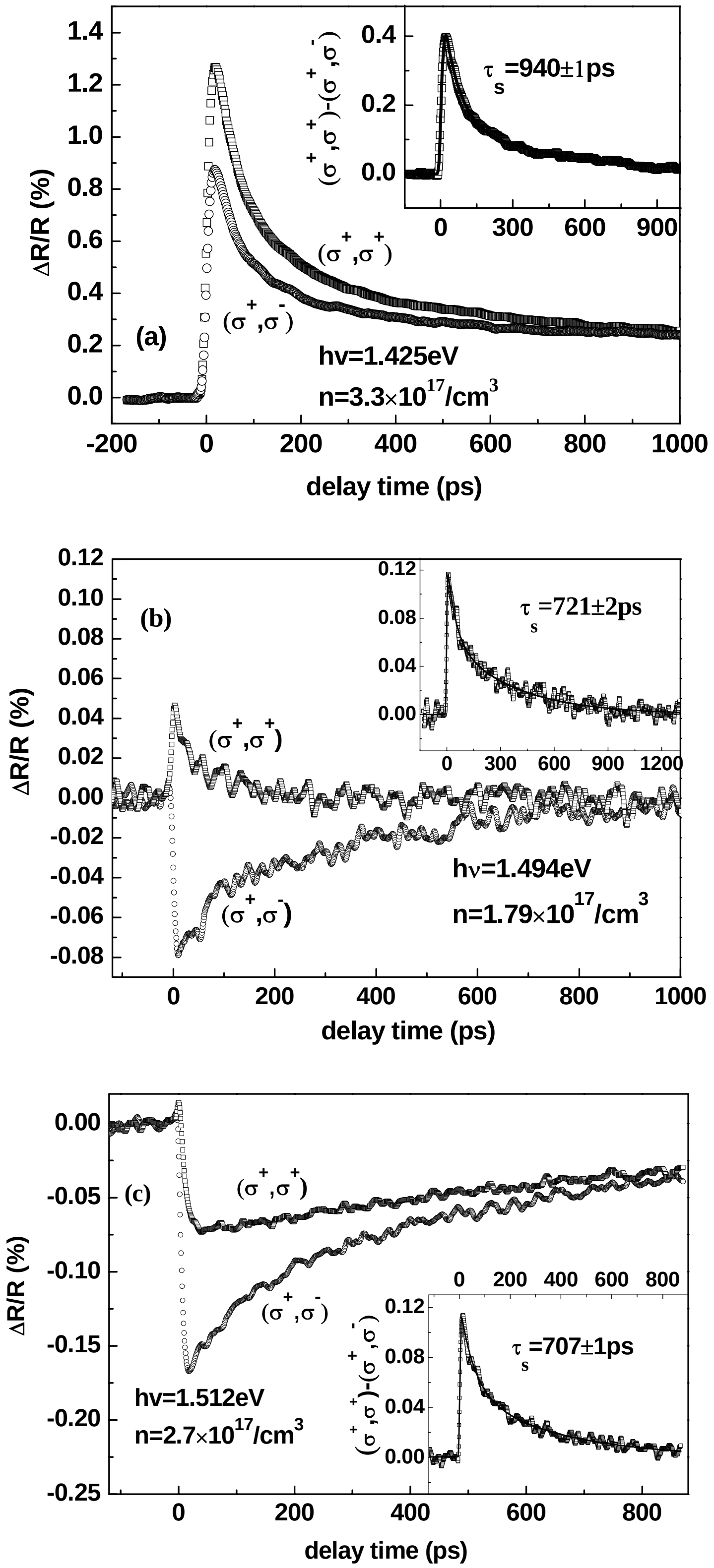

Fig. 2.

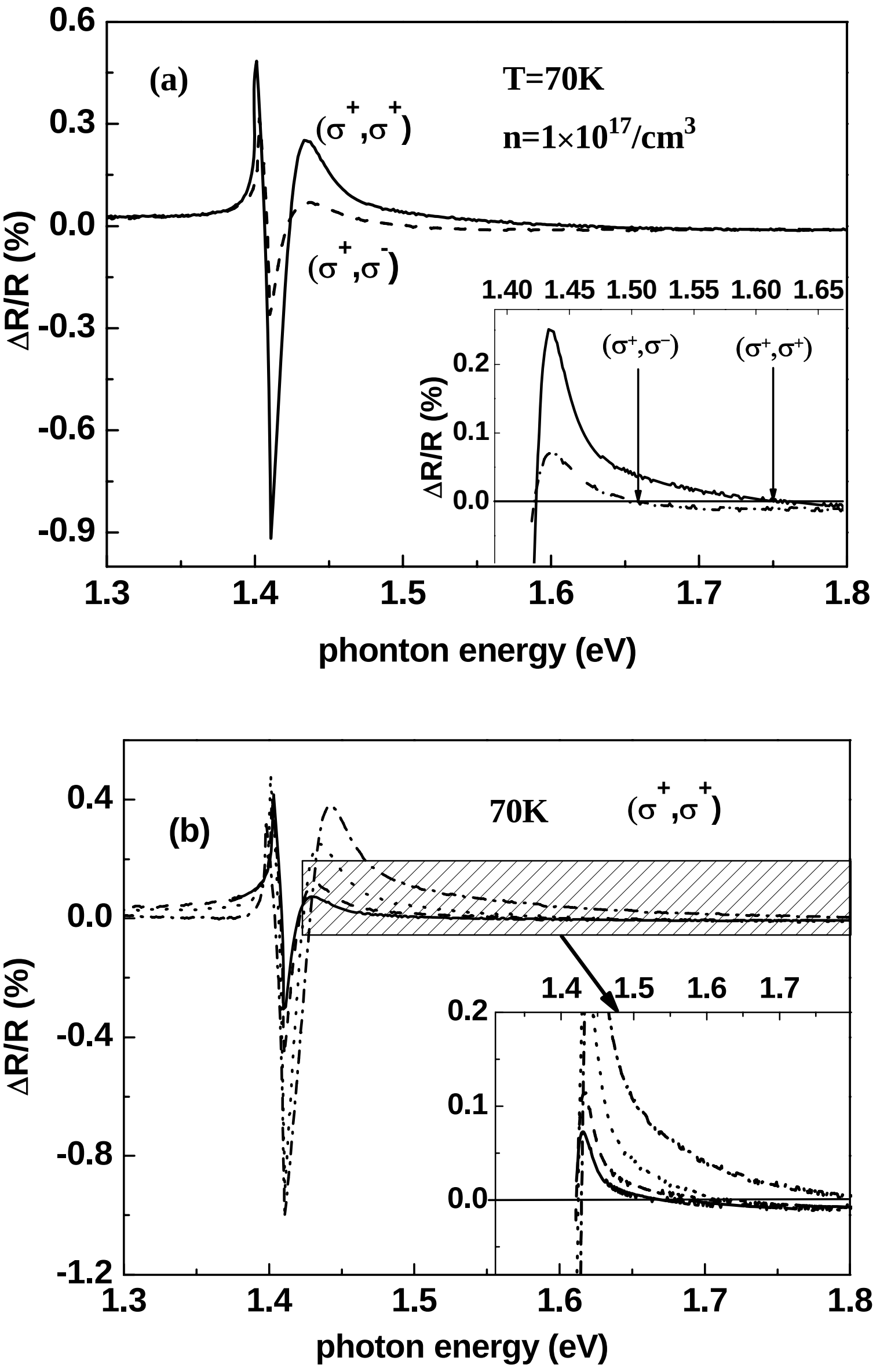

Fig. 3.

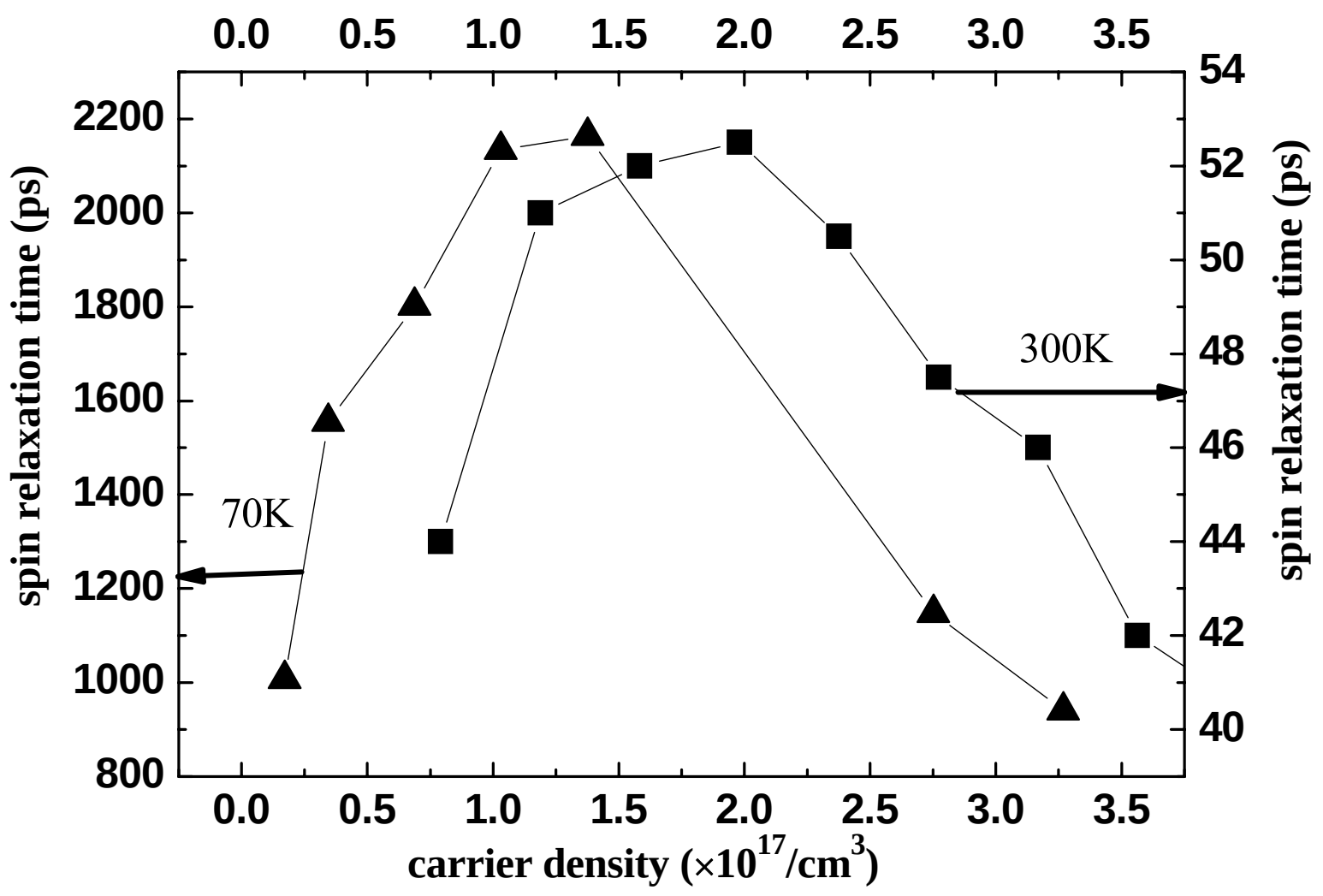

Fig. 4.